# Insights from Ex-Typhoon Halong (2025) – An Arctic Cyclone of Tropical Origin

Mingshi Yang [a], Zhuo Wang [a], John E. Walsh [b], James D. Doyle [c], Richard L. Thoman Jr. [b], Alice K. DuVivier [d]

[a] *University of Illinois Urbana-Champaign, Urbana, Illinois*

[b] *University of Alaska Fairbanks, Fairbanks, Alaska*

[c] *Naval Research Laboratory, Monterey, California*

[d] *National Science Foundation National Center for Atmospheric Research, Boulder, Colorado*

*Corresponding author*: Zhuo Wang, zhuowang@illinois.edu

ABSTRACT

An Arctic cyclone, Ex-Typhoon Halong, produced strong winds and devastating flooding in southwestern Alaska during 11–12 October 2025. This study examines the evolution of Halong after its transition into an extratropical cyclone through the analysis of ERA5 reanalysis and WRF model simulations. It is found that warm sea surface temperature (SST) anomalies over the western North Pacific preconditioned ex-Halong for intensification by increasing water-vapor content and reducing static stability. Quasi-geostrophic lifting associated with a subsequent interaction with another extratropical cyclone led to the rapid deepening of ex-Halong. This case demonstrates that tropical cyclones can transition into extratropical systems that are intensified by anomalously warm ocean waters, exacerbating impacts in high latitudes. Further analyses indicate that an increasing fraction of Alaskan cyclones has originated in tropical latitudes (south of 30°N) in recent decades. In particular, the frequency of Arctic cyclones of tropical origin increased by a factor of four in August and by a factor of three in September during 1980–2025 compared with 1940–1979.

SIGNIFICANCE STATEMENT

This study investigates the mechanisms behind the severe impacts of ex-Typhoon Halong on Alaska in October 2025. Our analysis demonstrates that unusually warm sea surface temperatures preconditioned the re-intensification of the storm in the extratropics. This intensification was further facilitated by the storm's interaction with an existing extratropical weather system, leading to rapid deepening. Beyond this specific event, the frequency of Arctic cyclones of tropical origins has increased four-fold during late summer since 1980. These findings suggest that as ocean temperatures continue to rise, tropical-origin storms may increasingly pose a major risk to high-latitude communities and environments.

CAPSULE

The re-intensification of ex-Typhoon Halong south of the Bering Sea is investigated, and we reveal a four-fold increase in late-summer Arctic cyclones of tropical origin over the North Pacific.

## 1. Introduction

During 11–12 October 2025, ex-Typhoon Halong produced devastating winds and storm-surge flooding in communities along Alaska's southwestern coast. The native villages of Kipnuk and Kwigillingok (Fig. 1) were especially hard hit, with a wind-driven storm surge that lifted houses off their foundations, caused water system outages, washed out roads, and damaged airfield runways. More than 1500 residents had to be airlifted (some by helicopter) to larger Alaskan towns and cities, where they would need to over-winter before returning to their damaged homes. In addition, flooding and prolonged power outages associated with ex-Halong led to substantial losses of stored subsistence foods and traditional harvests, worsening food security in affected communities beyond the immediate economic costs (Rosen 2025). Arctic cyclones originating in midlatitudes and the Arctic are receiving increased attention in the climate literature (e.g., Ulbrich et al. 2009; Catto et al. 2019; Sinclair et al. 2020; Wang et al. 2024; Yang et al. 2024). However, ex-Halong and several other recent storms show that tropical cyclones can transition to Arctic cyclones and have major impacts in high latitudes. Such Arctic cyclones of tropical origin have been studied less, and this storm event highlights the need for better understanding and forecasts of these types of high-latitude storms.

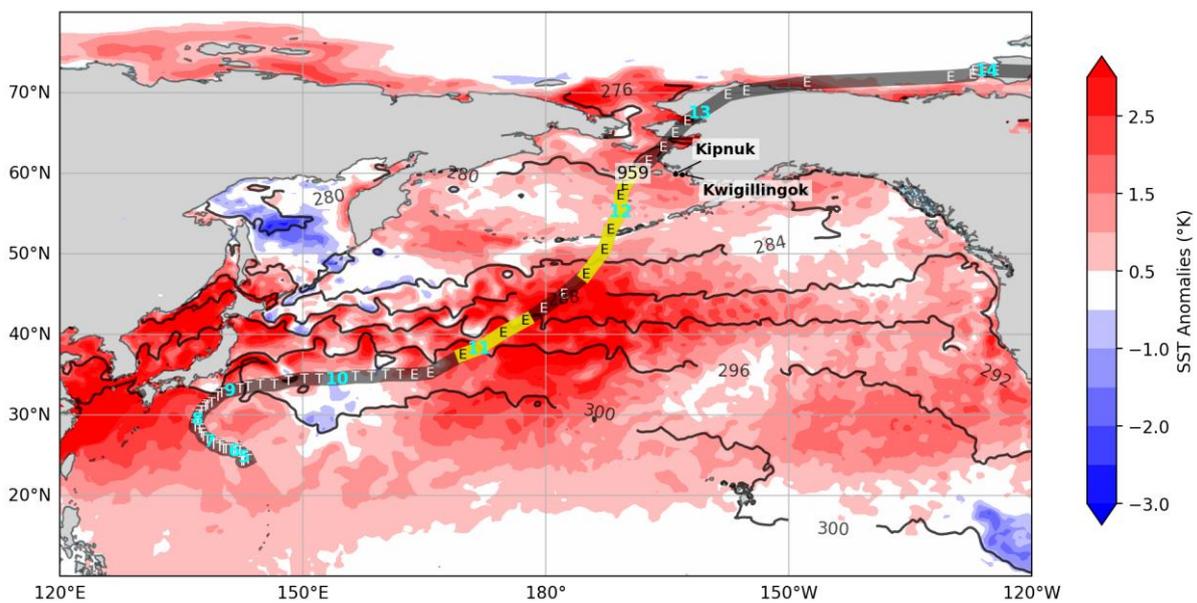

Figure 1. Track of Ex-Halong in ERA5 (gray line) superimposed on the long-term mean SST (black contours in K) and SST anomalies relative to the long-term mean (shading in K), averaged from October 4 to October 15, 2025. The location of Ex-Halong is marked by "T" during its typhoon stage, "E" after its extratropical transition, and light blue numbers during 0000 UTC of each date. Yellow segments represent two re-intensification phases of ex-Halong. "959" hPa marks the ex-Halong's peak intensity location at 0900 UTC on 12 October. Locations of Kipnuk and Kwigillingok villages are annotated in the map.

Storms in northern high latitudes are most common over the subpolar oceans of the North Atlantic and the North Pacific. However, storm activity in the North Pacific shows notable differences from the North Atlantic, mainly because of their different geographies. The North Pacific's ocean connection to the Arctic is via the Bering Strait, which is only about 50 km wide. Alaska and far-eastern Russia, both of which have major topographic features, serve as barriers between the North Pacific and the Arctic. By contrast, the oceanic connection between the North Atlantic and the Arctic extends for thousands of kilometers from the eastern Greenland coast to the Barents Sea. To the west of Greenland, Baffin Bay and the Labrador Sea have a narrow ocean connection to the Arctic via Nares Strait. High-latitude storm activity, which is most active in autumn and winter, is shaped by these geographical differences. Over the North Atlantic, the storm track is oriented southwest-northeastward, and many storms move from the North Atlantic to the Arctic (Yang et al. 2024). The North Pacific storm track, by contrast, is generally oriented west-to-east with no extension into the Arctic. Consequently, only a small fraction of North Pacific storms reaches the Arctic by passing northward through the Bering Strait.

The most comprehensive assessment of North Pacific cyclone characteristics was produced more than 15 years ago by Mesquita et al. (2010), whose summary statistics illustrate the generally zonal nature of the storm tracks from the western North Pacific to the Gulf of Alaska, a modest north-south seasonal shift in the storm track, and the role of barotropic processes in the Gulf of Alaska contributing to cyclolysis via shear destruction of transient eddies. Additionally, topographic blocking by the major Alaska mountain ranges also contributes to a center of lysis in the Gulf of Alaska. Interestingly, Mesquita et al.'s analysis shows that high storm intensities occur over two primary locations: south of the Bering Sea and in the Gulf of Alaska. Intensification in the Gulf of Alaska is especially apparent when referenced to the background climatology (Mesquita et al. 2010, their Fig. 14).

Through their combination of wind-driven waves and storm surge, storms are often the most consequential environmental events in coastal regions globally. Coastal flooding and erosion are especially significant hazards in high-latitude coastal areas where waves and flooding accelerate the thaw of ice-rich coastal permafrost. Coastal erosion rates in the Arctic are among the largest on Earth, with average rates of retreat of several meters per year along much of the coasts of Russia and Alaska (Irrgang et al. 2022), and rates exceeding 5 meters per year have been documented along parts of Alaska's Beaufort Sea coast (Gibbs and Richmond 2015). Various studies point to a doubling (and even more) of coastal erosion rates in the Arctic in recent decades (Arp et al. 2010; Overeem et al. 2011; Fredrick et al. 2016). While flooding and erosion are highly visible impacts, coastal storms also lead to saltwater intrusions that contaminate village water supplies. Even in areas not prone to coastal flooding, strong winds associated with storms can damage infrastructure and pose serious risks to transportation, commerce, and personal safety.

The impacts of changes in Arctic storm activity are compounded by changes in sea ice, which serves as a buffer protecting coastlines from wave-driven flooding and erosion. In this context, the U.S. Global Change Research Program has used the Alaska coastline to illustrate the increased risks of flooding and erosion when sea ice cover is not present (Karl et al. 2009). Since the open water season offshore of many Arctic coasts has lengthened by 1 to 3 months in recent decades (Parkinson 2022), storms across the Arctic pose increasing risks regardless of whether storm activity is changing. Occurring in this recent period of diminished sea ice cover, Halong as well as ex-typhoon Merbok in September 2022 provide physically plausible previews of the impacts of future Alaskan coastal storms, particularly as continued sea ice loss is expected to increase the exposure of northern coastal regions. Barnhart et al. (2014) showed that the retreat of sea ice has increased the open-water fetch for autumn storms along much of the Arctic Ocean's coastline and illustrate links between increased fetch and extreme values of water-level setup (their Figure 9). Rolph et al. (2018) showed that, over the period 1979-2014, there was nearly a tripling of the number of wind events during open water conditions at Utqiaġvik. Most of this increase was attributable to the increased open water season length, although the frequency of storm-related high-wind events has increased in this region (e.g., Rolph et al. 2018; Redilla et al. 2019). Kettle et al. (2025) have recently documented an increase in storm impacts at Nome, Alaska in the context of reduction in coastal sea ice. These compounding environmental changes exacerbate the potential impacts of a storm like Halong on the communities.

Halong was unusual in several respects. First, it intensified by more than 10 hPa in the Bering Sea before an unexpected eastward shift in its track. The storm then moved northeastward close to Alaska's southwest coast, where it weakened before entering the southern Beaufort Sea. In the 21$^{st}$ century, Halong is one of only three ex-typhoons to terminate north of the Bering Strait (the others are ex-typhoon Ampil in 2024 and ex-typhoon Merbok in 2022). Such tracks are unusual for any extratropical cyclone (e.g., Mesquita et al. 2010, their Fig. 14).

We will address the following questions in this study: What processes or factors contributed to the extratropical intensification of ex-Halong? How does ex-Halong compare to other Arctic cyclones of tropical origin? The remainder of the manuscript is organized as follows. Section 2 describes the data and methods. Section 3 provides a synoptic overview of Ex-Halong. Sections 4 and 5 investigate the roles of upper-level atmospheric processes and sea surface temperature (SST) anomalies through quasi-geostrophic (QG) diagnosis and WRF model simulations, respectively. Section 6 briefly examines the track biases of Ex-Halong forecasts. Section 7 presents the statistics of North Pacific cyclones in the Bering Sea sector, particularly cyclones of tropical origin, and Section 8 follows with a summary and discussion.

## 2. Data and Methods

*a. ERA5 reanalysis*

ERA5 (Hersbach et al. 2020) is a state-of-the-art, high-resolution global atmospheric reanalysis available from 1940 to the present, with updates typically released within about one week. The three-hourly ERA5 fields are used to track ex-Halong and other North Pacific and Arctic cyclones, analyze the cyclone structure and evolution of ex-Halong, and provide initial and boundary conditions for WRF simulations.

*b. Cyclone tracking*

Cyclone tracking is carried out using the algorithm of Crawford et al. (2021), applied to 3-hourly ERA5 sea-level pressure (SLP) from 1 January 1940 to 16 October 2025. The resulting dataset includes all Northern Hemisphere cyclones, spanning tropical, extratropical, and Arctic cyclones. A minimum lifetime of 24 hours and a minimum propagation distance of 1000 km are applied to filter out spurious transient and stationary systems, following previous studies (e.g., Crawford et al. 2021; Yang et al. 2024).

We examine two overlapping cyclone populations in the warm season (May to October) over the Pacific sector (i.e., cyclone must stay within the longitude range of 90°-255°E unless it reaches poleward of 65°N). First, we examine North Pacific Cyclones (NPCs) with potential impacts on the Alaska–Bering Strait region, defined by cyclones that travel in the region of (170°-200°E, 55°-65°N) for at least one timestep during its lifetime. These storms may form locally or originate from surrounding or remote, lower-latitude regions. Second, we identify a group of Arctic Cyclones of Tropical origin (ACTs). A cyclone is classified as an ACT if it undergoes genesis in the tropics (latitude <30°N) and later reaches the Arctic (latitude ≥65°N), following a rare northward pathway of cyclone propagation in the Pacific sector. Ex-Halong satisfies both criteria and serves as an example of how remote tropical systems can transition into exceptionally intense, high-impact storms at high latitudes.

*c. Composite analysis*

Following the methodology of Stoll et al. (2021) and Yang et al. (2024), storm centric composites are constructed to facilitate comparison of cyclones in different latitudes. ERA5 fields are projected from the native latitude-longitude grid to a 5,000 × 5,000 km square grid centered on the cyclone center (determined based on SLP minima) with the horizontal resolution of 25 × 25 km. The northern and southern boundaries of the square are oriented perpendicular to the longitude line passing through the cyclone center, while the eastern and western boundaries are parallel to that central longitude line. As a result, the zonal and meridional directions in the composites do not correspond exactly to the geographic east–west and north–south directions. For convenience, however, we still refer to them as east–west and north–south directions.

*d. Quasi-geostrophic diagnoses of vertical motion*

The Q vector form of quasi-geostrophic (QG) omega equations (Bluestein 1992; Holton 2004) is used to diagnose various processes contributing to cyclone evolution,

$$\left(\nabla_p^2 + \frac{f_0^2}{\sigma}\frac{\partial^2}{\partial p^2}\right)\omega = -2\nabla_p \cdot \boldsymbol{Q} + \frac{f_0\beta}{\sigma}\frac{\partial v_g}{\partial p} - \frac{\kappa}{\sigma p}\nabla_p^2 J, \qquad (1)$$

where

$$\boldsymbol{Q} \equiv -\frac{R}{\sigma p}\left(\frac{\partial \boldsymbol{V_g}}{\partial x}\cdot \nabla_p T, \frac{\partial \boldsymbol{V_g}}{\partial y}\cdot \nabla_p T\right). \qquad (2)$$

$\omega$ is vertical velocity, $\boldsymbol{V_g} = (u_g, v_g)$ is the geostrophic wind vector, $J$ is diabatic heating rate per unit mass, and $T$ is temperature. On the right-hand side (RHS) of Eq. 1, the first term corresponds to adiabatic QG forcing and the third term represents the role of diabatic heating (DH). The second term is related to the beta-effect and is neglected because its magnitude is much smaller than the other terms.

Laplacian inversion can be applied to Eq. 1 to derive the QG vertical motion $\omega$. Since the horizontal component of the Laplacian dominates the total Laplacian term on the LHS of Eq. 1 throughout ex-Halong's re-intensification (Fig. S1), we drop the vertical Laplacian term and solve for $\omega_F$ at each pressure level independently using the simplified two-dimensional Poisson equation

$$\nabla_p^2 \omega_F = F \qquad (3)$$

where $\omega_F$ represents the vertical motion associated with a forcing term $F$ (i.e., each or the sum of the first and third RHS terms in Eq. 1). As shown later in Section 4, the derived $\omega$ closely resembles that from the ERA5, justifying the simplification of Eq. 1.

Equation 3 was discretized on the storm centric square grid and solved using sparse direct solvers under a Dirichlet boundary condition, which assumes $\omega = 0$ at domain boundaries. A comparison between the perfect boundary condition (i.e., setting the boundary values using $\omega$ from ERA5) and the Dirichlet boundary condition shows that the differences between reconstructed omega fields are primarily confined near the boundaries (Figure S2). Since the diameter of ex-Halong is smaller than the domain size (i.e., 5,000 × 5,000 km), the Dirichlet boundary condition is applied for simplicity.

e. Numerical model experiments

To investigate how SST influences the re-intensification of ex-Halong, we conduct a set of sensitivity experiments using the Weather Research and Forecasting (WRF) model version 4.2.2 (Skamarock et al., 2019). The simulations run from 0000 UTC 10 October to 0000 UTC 13 October, covering the period during which ex-Halong re-intensifies. The model domain spans 8,400 km in both horizontal directions with a grid spacing of 20 km, ensuring that the storm center is at least 2,000 km away from the domain boundaries.

Initial and three-hourly boundary conditions are taken from the ERA5 reanalysis. Different time-invariant SST configurations are applied in three experiments. The control experiment (CTRL) uses the original ERA5 SST field at 0000 UTC 10 October 2025. In the second experiment (LTM), the long-term mean SST at 0000 UTC 10 October averaged over 1940-2025 is used to drive the WRF model. In the third experiment (DSST), the linear trend of the SST during 1940-2025 is removed at each grid point, and the detrended SST at 0000 UTC 10 October 2025 is used (see Figs. 6a and b and more discussion in section 5). For each experiment, 18 ensemble simulations are performed by perturbing the physical parameterizations (Table S1). Because Halong propagates from the tropics into the Arctic, physics options are sampled from all valid combinations of the NCAR convection-permitting and tropical physics suites (NCAR/UCAR 2026), supplemented by additional configurations recommended for Arctic applications (Ban et al. 2023).

An addition set of experiments with different initialization times are carried out to assess ex-Halong's track predictability. More information about these experiments is provided in Section 6.

## 3. Synoptic Evolution of Ex-Halong

The track of Halong begins at 0600 UTC 4 October over the western North Pacific (24.5°N, 143.25°E) as identified by the cyclone tracker using the ERA5 reanalysis. SSTs in the genesis region were higher than 300 K, exceeding the climatological mean by >1 K (Fig. 1). Initially, Halong tracked northeastward, reaching 30°N on 8 October, and then turned eastward. While ERA5 accurately resolves the track of Typhoon Halong, closely aligned with the typhoon track from the Japan Meteorological Agency (JMA 2026) analyses, it underestimates the storm intensity (Fig. 2a). The minimum sea level pressure (SLP) from the ERA5 is ~970 hPa, in contrast to 935 hPa estimated by JMA (Fig. 2a). Furthermore, the timing of peak intensity in ERA5 (2100 UTC 8 October) lags the JMA analysis (1800 UTC 7 October) by approximately one day (Fig. 2a).

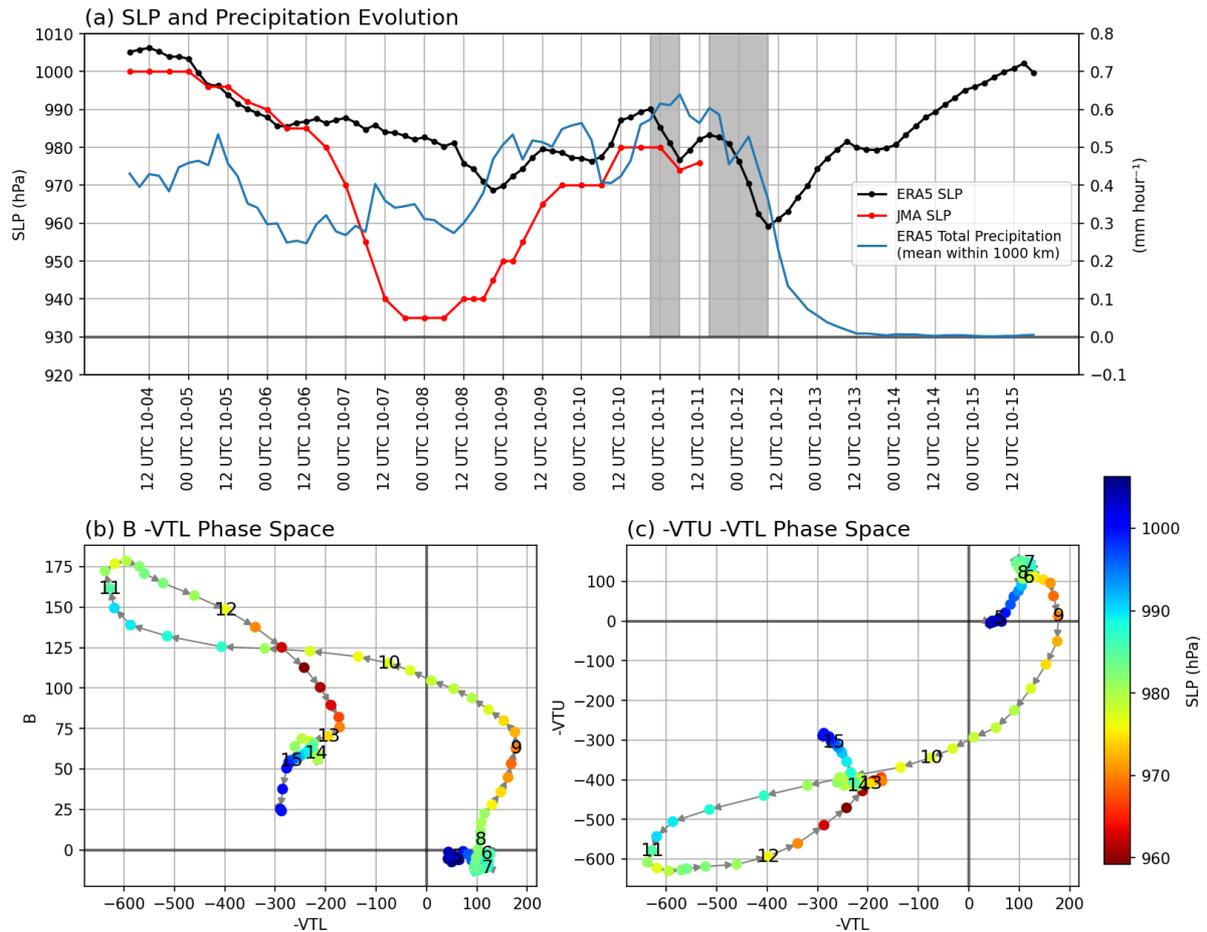

Figure 2. Top: time series of the minimum SLP (in hPa; ERA5 in black, and JMA in red until 1200 UTC 11 October) and the total precipitation (blue; averaged within 1000 km radius of the cyclone center; units: mm hr$^{-1}$). Shaded areas represent two re-intensification phases of ex-Halong in ERA5. Bottom: cyclone phase space diagrams from 0600 UTC 4 October to 0018 UTC 15 October, derived using ERA5, with numbers indicating 0000 UTC of each date. The metric B represents thickness asymmetry following the moving storm, and the metrics -VTL and -VTU represent the cyclone thermal structure in the lower troposphere (900-600 hPa) and upper troposphere (600-300 hPa), respectively, with positive (negative) values for a warm-core (cold-core) structure.

Halong underwent extratropical transition (ET) on 10 October (Fig. 1). The ET of Halong is illustrated by the Cyclone phase space analysis (Hart 2003) (Figs. 2b-c). Halong evolved from a symmetric, deep warm-core tropical cyclone before 8 October into a highly asymmetric, deep cold-core system after 10 October. The thermal asymmetry and cold core were strongest on 11 October and gradually decreased afterwards as the storm weakened. While JMA ceased tracking the storm soon after the completion of ET, ex-Halong in ERA5 accelerated northeastward after 10 October, nearly tripling its propagation speed from that before ET. During this phase, Ex-Halong traversed over a region of warm SST anomalies (>3K) centered near 40°N, 180°E, and it re-intensified from 990 hPa at 1800 UTC 10 October to a peak intensity of 959 hPa at 0900 UTC 12 October over the North Pacific (60°N, 169.25°W). Ex-Halong subsequently made landfall and crossed western Alaska while weakening, and later dissipated over the Canadian Arctic after 14 October.

During the re-intensification stage, two distinct episodes of deepening with SLP change rate stronger than 1.5 hPa h$^{-1}$ appeared in the ERA5 track (shaded areas in Fig. 2a). The first occurred from 2100 UTC 10 October to 0600 UTC 11 October as Ex-Halong moved across the region of warm SST anomalies exceeding 3 K (Fig. 1). A second, more persistent deepening episode occurred from 1500UTC 11 October to 0900 UTC 12 October, leading to the storm's peak intensity. During this second deepening episode, SST fell below 288 K, and the magnitude of warm SST anomalies decreased to roughly 1 K. Precipitation associated with the cyclone peaked right after the first intensification episode (Fig. 2a) and then decreased by about 50% by landfall at 1200 UTC 12 October. While there was still heavy precipitation at landfall, the storm surge driven by strong winds was the major contributor to the significant coastal flooding in western Alaska. Additionally, enhanced soil moisture from preceding precipitation also contributed to the flooding.

Cyclone centric composite snapshots from 0000 UTC 10 October to 1200 UTC 12 October are shown in Figure 3. The northeastward acceleration of ex-Halong after 1800 UTC 10 October and during 11 October was facilitated by a large, quasi-stationary extratropical cyclone (ETC) centered just east of Kamchatka Peninsula (Figs. 3a, 3g). The ETC featured a deep circulation and steered ex-Halong cyclonically around it. Ex-Halong developed a pronounced frontal structure during 11-12 October, with persistent precipitation northeast of the center and a frontal boundary to the south, forming a comma-shaped precipitation pattern (Figs. 3e-f; k-l). Strong upward surface heat fluxes occurred northwest of ex-Halong during 10-11 October, largely associated with the cold northerly flow southwest of the ETC and potentially enhanced by positive SST anomalies. The column integrated water vapor fluxes indicate moisture transport from this region toward ex-Halong (Figs. 3g-3i), helping to fuel its precipitation.

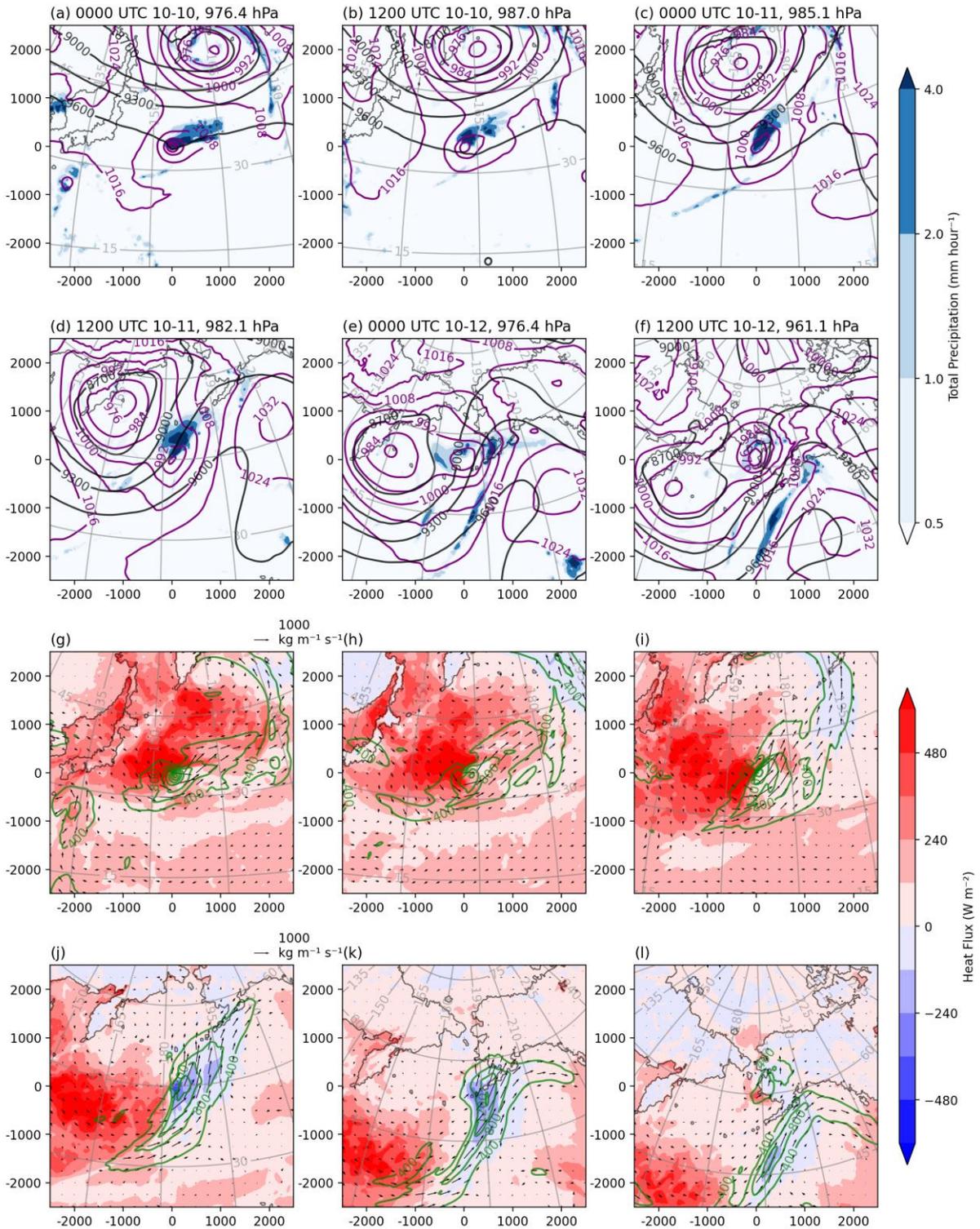

Figure 3. Top two rows: SLP (purple contours in hPa), H300 (black contours in m), and total precipitation (shading in mm hour$^{-1}$); Bottom two rows: vertically integrated water vapor flux (arrows in kg m$^{-1}$ s$^{-1}$) and its magnitude (green contours in kg m$^{-1}$ s$^{-1}$) from 1000 to 1 hPa and upward net heat flux (shading in W m$^{-2}$). Titles in the top rows indicate the timestamp and the minimum cyclone central pressure. Gray lines denote latitude, longitude lines every 15°. Axes represent the zonal (x) and meridional (y) distance from the cyclone center in km.

As ex-Halong approached Alaska (Figs. 3e-f, k-l), it reached its peak intensity at 0900 UTC 12 October and was accompanied by strong southerly moisture transport east of the cyclone center. The enhanced poleward moisture flux and associated high winds contributed to storm surge impacts, coastal flooding, and widespread precipitation across western Alaska. Backward trajectory analysis using the HYSPLIT model (Fig. S3) shows that moist air originated from the Central Pacific as well as southwest of the ETC. As Halong crossed the dateline at 0900 UTC 11 October, the near-surface southerly winds intensified further and accelerated the northward moisture transport into Alaska. The specific humidity of the parcels decreased rapidly on 12 October as they approached the coast, indicating moisture loss primarily due to heavy precipitation and a colder environment (Fig. S3).

In summary, the re-intensification of ex-Halong played a central role in its devastating impacts over western Alaska. Several processes may have contributed to its re-intensification. One factor is the passage of the storm over anomalously warm SSTs. The enhanced surface heat fluxes may have increased column water vapor and reduced static stability, preconditioning ex-Halong for its re-intensification. Additionally, Ex-Halong moved between the quasi-stationary ETC and the upper-level ridge to the east. These systems may have interacted with ex-Halong and contributed to its re-intensification. We will evaluate the possible contributions of these processes in Sections 4 and 5.

## 4. Role of Upper-Level Forcing and Diabatic Heating in the Re-intensification of Ex-Halong

We first investigate the role of the upper-level forcing in the re-intensification of ex-Halong. The vertical motion diagnosed from upper-level adiabatic QG forcing and diabatic heating (via Eqs. 5-7) is evaluated against omega from ERA5 (Fig. 4). Two timesteps are shown: the first row shows 0000 UTC October 11, corresponding to the first intensification episode, and the second row shows 0600 UTC October 12, representing the second intensification episode. During both times, strong ascents are evident above and downstream of the surface cyclone with descending motion upstream.

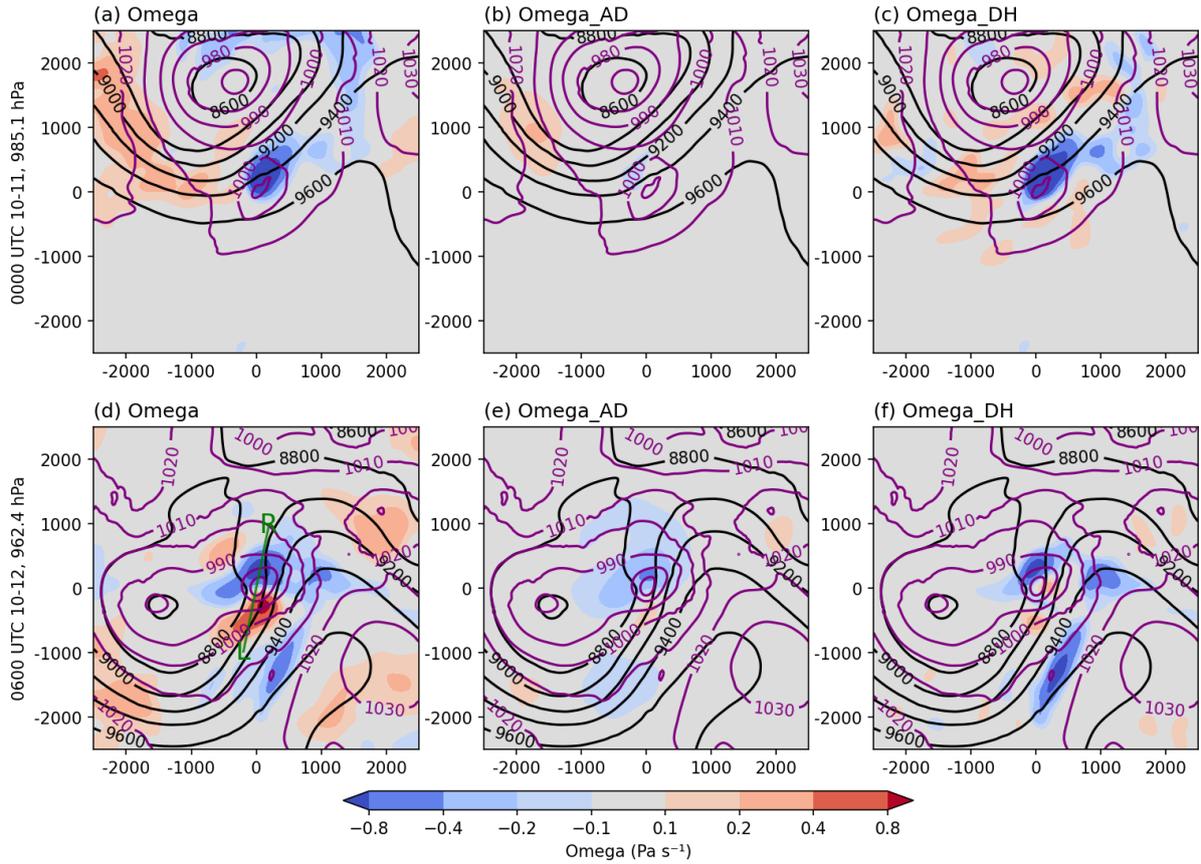

Figure 4. Omega from ERA5 (left column) and derived from adiabatic QG forcing (Omega_AD; middle column) and diabatic heating (Omega_DH; right column) averaged vertically between 600 and 300 hPa. H300 (m) and SLP (hPa) are shown in black and purple contours, respectively. A Gaussian kernel smoothing (sigma=100 km; full width at half maximum ~ 235 km) is applied to the diagnosed omega fields to highlight synoptic-scale features. Axes represent the zonal (x) and meridional (y) distances from the cyclone center in km. The cross section for Figure 5 is indicated in panel (d), where "L" and "R" denote left and right extents of Fig. 5.

Omega induced by diabatic heating accounts for most of the localized and intense ascent during the first intensification period while adiabatic QG forcing is minimal (Figs. 4a-c), and the spatial distribution of diabatic heating-driven ascent aligns well with heavy precipitation northeast of Ex-Halong (Fig. 3c). During the second intensification phase, the adiabatic QG forcing is largely confined along the upper-level jet (indicated by the strong gradient of H300) between the ETC and the ridge (Fig. 4e). It is broader in horizontal scale and weaker in magnitude compared with omega induced by diabatic heating, which primarily enhances fine-scale ascent features (Fig. 4f). This suggests that adiabatic QG forcing associated with the upper-level jet may have promoted vertical motion, which is further amplified by diabatic heating. Meanwhile, it should be noted that the limitations of the omega inversion method, which neglects vertical coupling, may underestimate the reconstructed vertical velocity. Taken together, while the first diabatic-heating-driven intensification phase is weak and transient, the subsequent interaction between the upper-level systems and ex-Halong, aided by diabatic processes, leads to a more sustained and rapid intensification. The adiabatic QG-driven ascent promotes precipitations over and ahead of the surface cyclone center, and the resultant latent heat release in turn reinforces localized upward motion. Further analysis shows that the adiabatic QG forcing is largely confined to the upper troposphere while the diabatic QG forcing is pronounced in both the upper and lower troposphere (Figs. S1 and S5).

To better understand the upper-level QG forcing, a vertical cross section (Fig. 5) is constructed across the surface cyclone center, extending from southwest to northeast (see the cross section in Fig. 4d). Two vorticity maxima are evident along the cross section: one below 600 hPa with the maximum near the surface cyclone center, and another between 500 and 300 hPa with the maximum around 400 hPa slightly southwest of the surface cyclone center. The two centers correspond to the surface and upper-level circulations of ex-Halong. The upper-level jet exceeds 80 m s$^{-1}$ near 350 hPa, peaking upstream of ex-Halong. A pronounced dipole in QG forcing is located above the surface cyclone center. Q vector convergence (contributing to ascent) dominates downstream (i.e., northeast) and ahead of the surface cyclone center of ex-Halong, extending from the upper troposphere to near the surface, while Q vector divergence (contributing to descent) appears upstream (i.e., southwest) of the cyclone center. This pattern is consistent with vorticity advection by the thermal wind: the stronger upper-level jet implies a northeastward thermal wind, and it produces cyclonic vorticity advection northeast of the cyclone center and anticyclonic vorticity advection to the southwest. As a result, adiabatic QG forcing favors an ascent ahead of the cyclone and a descent behind it, contributing to both the deepening and northeastward motion of ex-Halong.

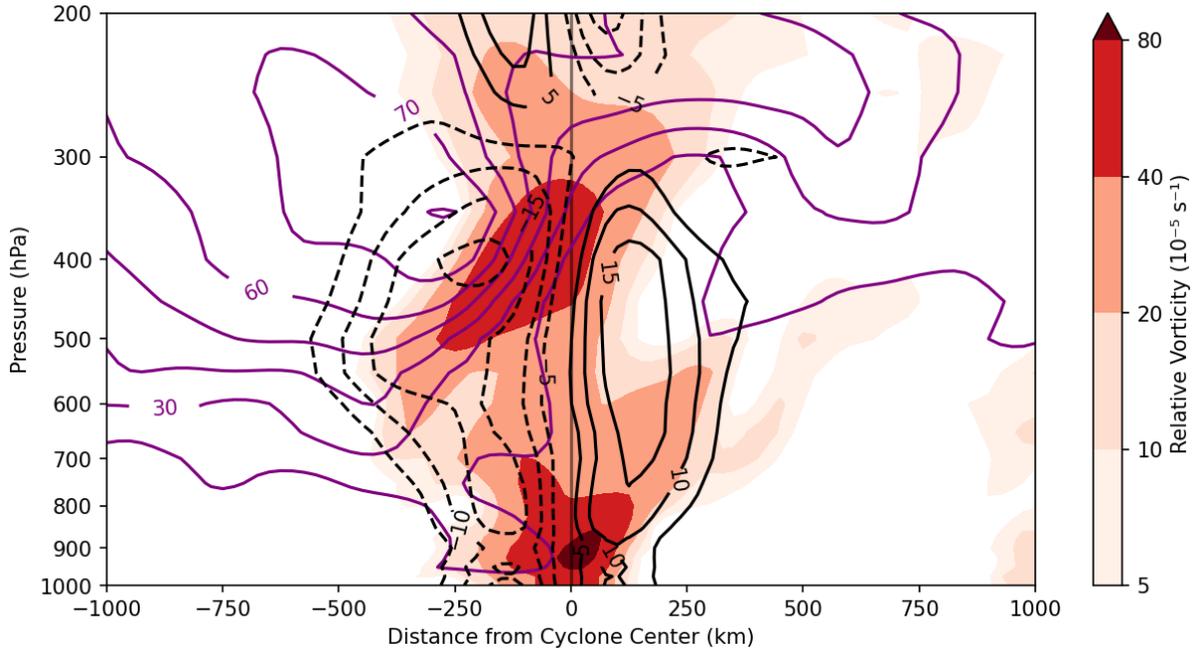

Figure 5. Vertical cross sections of relative vorticity (shading in $10^{-5}$ s$^{-1}$), adiabatic QG forcing (the first RHS term $-2\nabla_p \cdot \mathbf{Q}$ in Eq. 1; black contours every $5\times10^{-12}$ Pa s$^{-1}$ m$^{-2}$ with zero omitted), and tangential wind (from left to right along the cross section; purple contours every 10 m s$^{-1}$ starting from 20 m s$^{-1}$) at 0600 UTC 12 October. The horizontal axis shows the distance from the cyclone center (positive values in the northeast direction).

## 5. Role of SST in the Re-intensification of Ex-Halong in WRF Simulations

The preceding section highlights the influence of adiabatic QG forcing and latent heating on ex-Halong's re-intensification, with the latter potentially preconditioned by warm SST anomalies through enhanced column water vapor and reduced static stability. The role of warm SST anomalies and trend (Figs. 6a and 6b) is examined in this section using WRF model simulations.

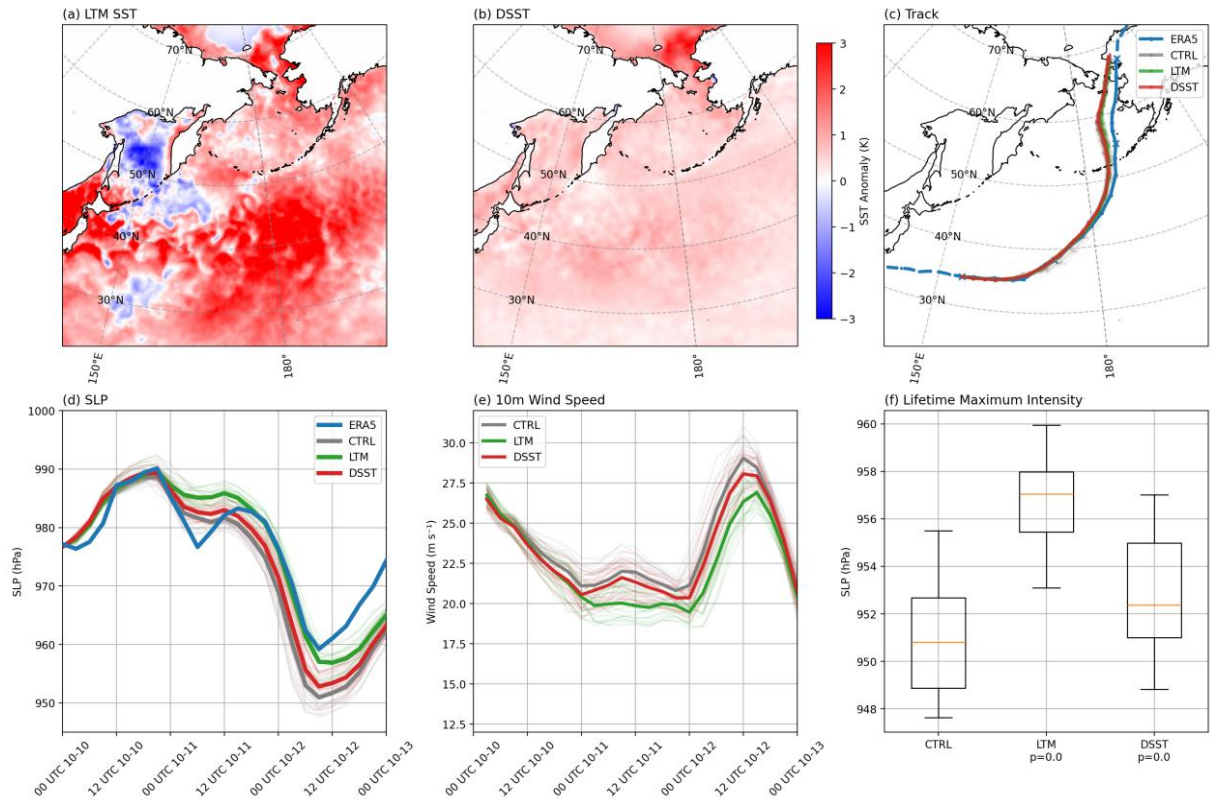

Figure 6. Top: SST differences (K) removed in the (a) LTM and (b) DSST experiments relative to CTRL, and (c) tracks of Halong in ERA5 and WRF simulations. Bottom: Time series of (d) the storm minimum sea level pressure and (e) 99$^{th}$ percentile 10-m wind speed (within 1000 km of the cyclone center), and (f) distributions of lifetime maximum intensity of ex-Halong in WRF experiments. Thick lines in (c-e) denote ERA5 or ensemble means, and thin lines denote individual ensemble members. P-values in (f) indicate the significance of differences relative to CTRL.

The track and intensity evolution from all experiments are shown in Figure 6. The cyclone tracks (Fig. 6c) do not show a strong spread between experiments or ensemble members, indicating the negligible influence of SST and physics configurations on ex-Halong's propagation during its re-intensification. This suggests the dominant role of the steering flow associated with the quasi-stationary ETC in governing Ex-Halong's trajectory. However, the cyclone tracks all exhibit a westward displacement over the Bering Strait compared to the observed track from ERA5. This track bias also occurred in operational forecasts until approximately 36 hours before Halong reached the coast. Such track biases have significant implications for impact forecasting over western Alaska and are briefly examined in the next section.

In terms of intensity evolution (Figs. 6d-e), the CTRL experiment underestimates the initial SLP deepening during the first intensification phase from 2100 UTC 10 October to 0900 UTC 11 October, but initiates the second intensification earlier at 1200 UTC 11 October, reaching an ensemble mean minimum pressure of 951 hPa, lower than ERA5 by 8 hPa. Removal of the SST anomalies (i.e., the LTM experiment) reduces the intensification rate in both episodes and leads to a weaker ensemble mean with a higher ensemble mean minimum pressure of 957 hPa. In contrast, detrending the SST (i.e., the DSST experiment) only moderately reduces the intensification rate and peak intensity of ex-Halong, yielding an ensemble mean minimum pressure of 953 hPa. The wind speed evolutions in various experiments are shown in Figure 6e. Wind speed decreases in all experiments during Halong's ET on 10 October and shows little increase during 0000 UTC - 1200 UTC 11 October, corresponding to the first re-intensification phase. Notably, the ensemble-mean wind speed in LTM shows almost no increase during 11-12 October. Following the second re-intensification phase, wind speed rapidly increases during 0000 UTC - 1200 UTC 12 October, then, the CTRL experiments exhibit the strongest winds, with a 99th-percentile 10-m wind speed of ~29 m s$^{-1}$. Ensemble-mean wind speeds in the LTM and DSST experiments are lower than CTRL by about 2 m s$^{-1}$ and 1 m s$^{-1}$, respectively.

The ensemble statistics of SLP-based intensity of ex-Halong in different experiments are further illustrated in Fig. 6f. Statistical significance of the differences between the sensitivity experiments and CTRL is determined via a Student's t-test with the null hypothesis that the mean difference between paired ensemble members is zero. The peak intensity of ex-Halong in the LTM and DSST experiments is significantly weaker than that in the CTRL experiment, although the change of intensity in DSST is smaller than that in LTM. The sensitivity of the intensification rate to SST varies between the two re-intensification episodes. During the first intensification episode (Figs. 6d&e and Fig. S4a), the LTM experiment exhibits a significantly weaker deepening rate than CTRL, underscoring the importance of regional SST anomalies (Figs. 1). In contrast, the DSST experiment shows an insignificant reduction, likely due to the relatively weak long-term SST trend (Fig. 6b). During the second intensification episode (Figs. 6d&e and Fig. S4b), both SST modifications lead to significantly weaker deepening. Overall, the experiments suggest that both SST trend and SST anomalies contributed to the intensification of ex-Halong, and that the latter had a stronger influence, which may be due to the relatively weak SST trend (Fig. 6b). It is also worth noting that the second re-intensification episode occurred after the storm has passed the region of maximum SST anomalies. The remote or persistent impacts of warm SST anomalies on the storm intensification are consistent with Kuo et al.'s (1991) findings that surface energy fluxes within 24 hours before cyclone intensification strongly affect cyclone development. Other studies have also shown that moisture originating far ahead of a cyclone can contribute to cyclone precipitation (Boutle et al. 2010; Dacre et al. 2019).

The role of SST during ex-Halong's evolution is further examined in Fig. 7. With the removal of the warm SST anomalies (Fig. 7a), the LTM experiment produces a substantial reduction in surface turbulent heat fluxes (Fig. 7b) and precipitation rate (Fig. 7c). Vertical cross-sections of specific humidity and temperature following the cyclone center reveal an overall cooling and drying tendency in the troposphere in the CTRL experiment due to the poleward motion of the storm (Fig. 7d). Compared to the CTRL experiment, the LTM experiment shows stronger cooling and drying signals extending upward from the surface and peaking around 0000 UTC 11 October (Fig. 7e), consistent with the strong reduction in SST and surface fluxes around this time (Figs. 7a-b). This supports our hypothesis that warm SST anomalies precondition the storm's re-intensification by enhancing column water vapor and reducing static stability. In contrast, the changes in surface fluxes, precipitation, lower tropospheric temperature and humidity due to SST detrending are rather modest (Figs. 7b-c and 7f), consistent with the relatively weak SST trend (Fig. 6b).

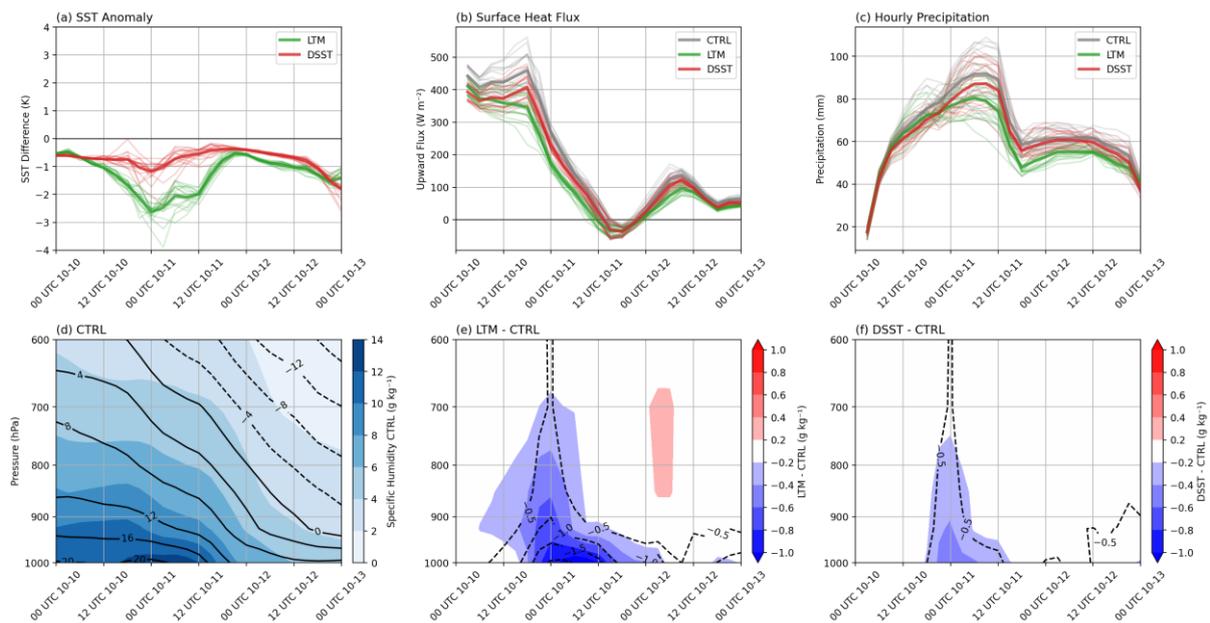

Figure 7. Top: (a) SST differences (K) averaged within 500 km of the cyclone center between sensitivity experiments and CTRL; (b) upward surface heat fluxes averaged within 500 km of the cyclone center in each experiment; and (c) the 99th percentile of hourly precipitation within a 1000 km radius of the cyclone center. Bottom: Time-pressure cross sections of ensemble-mean specific humidity (shading; g kg$^{-1}$) and air temperature (contours; K) averaged within 1000 km of the cyclone center, showing (d) CTRL, and differences between sensitivity experiments and CTRL for (e) LTM and (f) DSST.

## 6. Track Predictability of Ex-Halong

In the WRF simulations examined in the previous section, ex-Halong's trajectory starts deviating westward from the ERA5 track around 1200 UTC 11 October, coinciding with the cyclone's second re-intensification stage (Figs. 2 and 6c). This westward displacement exhibits little ensemble spread and shows limited sensitivity to the choice of model physics or SST boundary conditions, suggesting that, rather than intrinsic forecast uncertainty, the deviation reflects systematic aspects of the model configuration or its representation of the large-scale circulation representation. The latter could have resulted from reduced radiosonde launches, which may have negatively affected operational forecasts of ex-Halong (Sopow 2025). To better understand track biases, a series of additional experiments were performed with progressively later initialization times (Figs. 8a-b), all with a fixed NCAR Convection-Permitting Suite (NCAR/UCAR 2026; see the last entry in Table S1). The track error at the peak intensity time decreases with the decreasing forecast lead time as expected. However, it is worth noting that this improvement remains limited until the initialization time occurs after 1200 UTC 11 October. This timing immediately precedes the westward deviation of WRF-simulated ex-Halong in Fig. 6c. Because ex-Halong's northward propagation and subsequent re-intensification are likely influenced by its interaction with the nearby ETC, which is sensitive to storm intensity, we next examine the intensity biases of both systems in the CTRL ensemble simulations (initialized on 0000 UTC 10 October). As shown in Figures 6d and S6, the WRF simulations tend to overestimate the intensity of Halong and the ETC prior to the westward deviation. The overestimated cyclone intensities may have amplified the mutual attraction between two systems and produced the westward track biases of ex-Halong (Figs. 8a-b). This possibility is supported by Fig. 8c-d, which show that the intensity biases of the ETC and ex-Halong at 1200 UTC 11 October are significantly correlated with the track error at 0900 UTC 12 October. These results motivate further investigation into how intensity errors in interacting cyclone systems influence the predictability of ex-Halong's track.

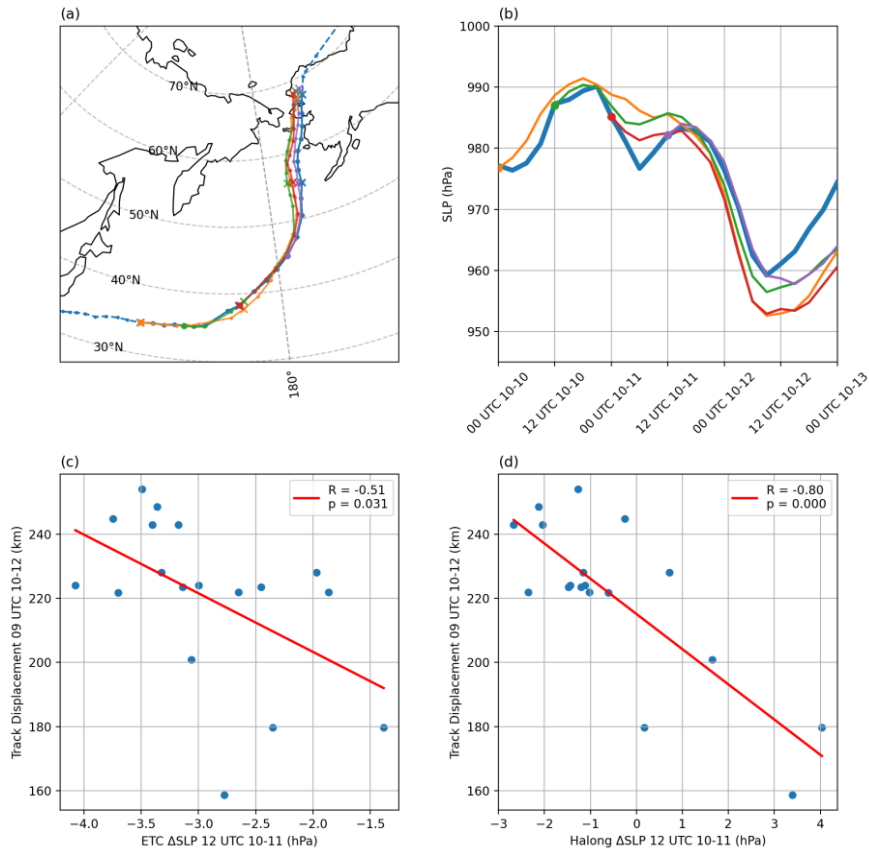

Figure 8. Top: tracks and SLP evolution from ERA5 (blue lines) and a set of WRF simulations initialized after 0000 UTC October 10 in 12-hour increments (shown in orange, green, red, and purple curves). 0000 UTC timesteps are marked by "X" in (a), and dots in (b) mark the initialization time for each simulation. Bottom: scatterplots with the linear fit between the intensity errors for (c) the quasi-stationary ETC and (d) Ex-Halong and track location errors at Ex-Halong's peak intensity time. Dots represent ensemble members in CTRL.

## 7. Climatology of North Pacific Cyclones and Arctic Cyclones of Tropical Origin

To place Halong within a broader climatic context, warm-season NPCs in the Bering Sea region (170°-200°E, 55°-65°N) are first examined (see bounding boxes in Fig. 9). The genesis density during 1940-2025 (Fig. 9a) shows that most cyclones form either within or close to the region, extending from Alaska southwestward towards Japan, consistent with the orientation of the Pacific storm track. Ex-Halong features a relatively low genesis latitude among NPCs. A comparison between the earlier decades (1940-1979) and recent decades (1980-2025) reveals several notable changes in genesis locations (Figs. 9b-d). In earlier decades (Fig. 9b), genesis was more tightly clustered southwest of the bounding box, reflecting a relatively confined region of NPC origin. In contrast, the recent period (Fig. 9c) exhibits a broader and more dispersed genesis density pattern. The difference field (Fig. 9d) highlights reduced genesis within and near the bounding box and enhanced genesis from more remote regions, including Northeast Asia, the subtropical western and central Pacific south of 40°N, and even poleward of the Bering Strait. A comparison of 200-hPa zonal wind during the warm season (Fig. 9d) between the early and recent decades reveals an enhanced westerly flow extending from Japan northeastward towards the Bering Strait in the recent decades, which may steer more cyclones from lower latitudes towards the Bering Strait and contribute to more NPCs of tropical origin in the recent decades.

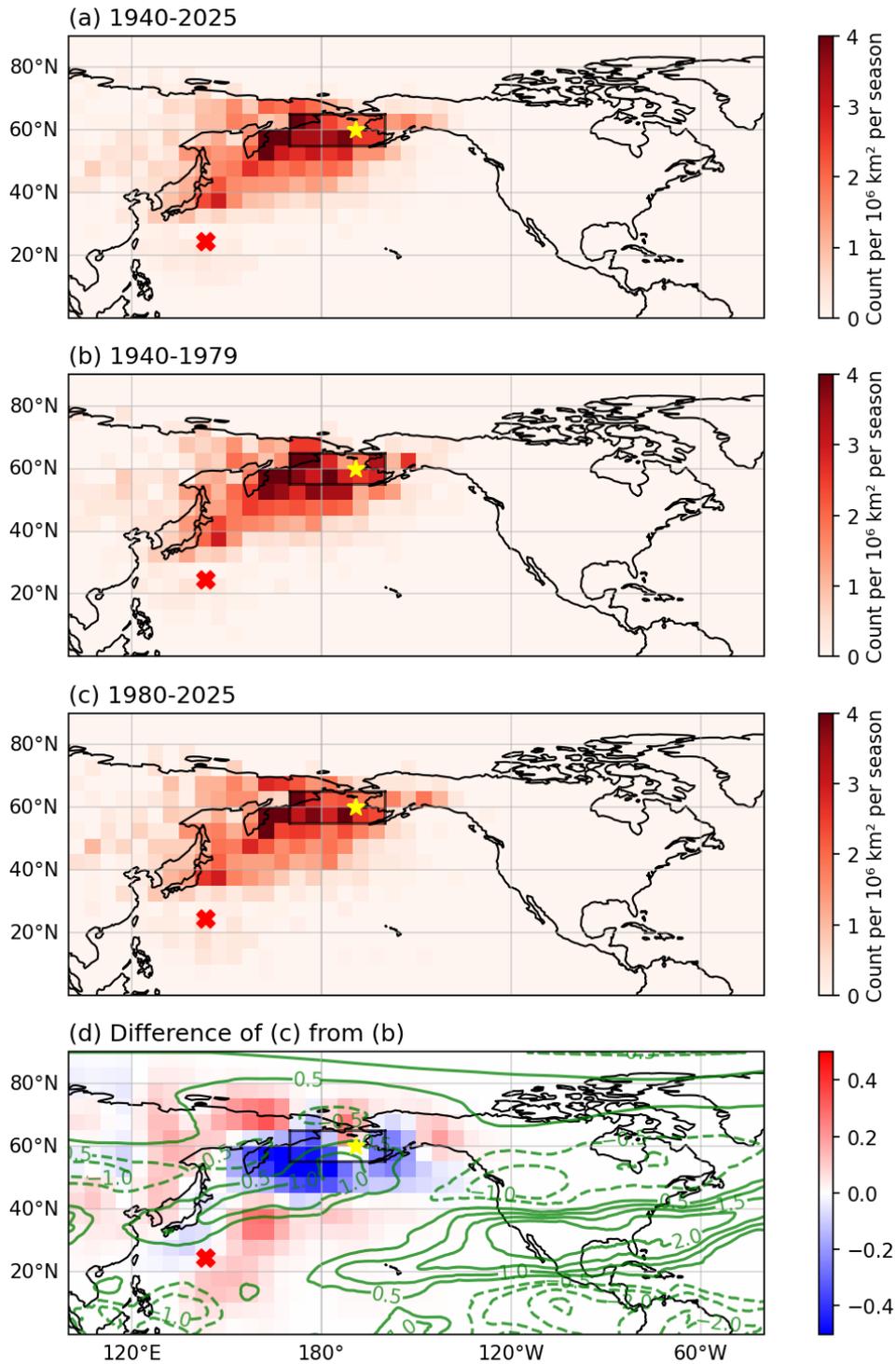

Figure 9. Genesis density function (shading) of warm-season NPCs in different time periods (a-c). The bounding box is (170°-200°E, 55°-65°N). Genesis and peak-intensity locations of ex-Halong are indicated by a red cross and a yellow star in each panel. (d) shows difference of genesis density function (shading; after gaussian smoothing of sigma=5°) and warm-season average of 200-hPa zonal wind (green contours in m s$^{-1}$) between two periods.

The moisture distribution of the NPCs when located within (170°-200°E, 55°-65°N) is examined in Fig. 10. On average, the cyclone-centric total column integrated water vapor (TCWV) fields exhibit a characteristic dry-west and moist-east asymmetry, with enhanced TCWV near the cyclone center (Fig. 10a). An overall increase in TCWV (>0.6 kg m$^{-2}$) has been found in the recent decades (Fig. 10b), especially at and east of the cyclone center. This may be attributed to a warmer atmosphere and more cyclones originating from lower latitudes. Ex-Halong is characterized by a distinct moist tongue east of the cyclone center, resembling a warm conveyor belt or an atmospheric river. TCWV in the moist tongue reaches 50 kg m$^{-2}$, exceeding the 90th percentile of the sample distribution and nearly doubling the composite mean (Fig. 10c). The high TCWV and the associated precipitation contribute to the strong intensity and high impacts of ex-Halong as discussed in a previous section.

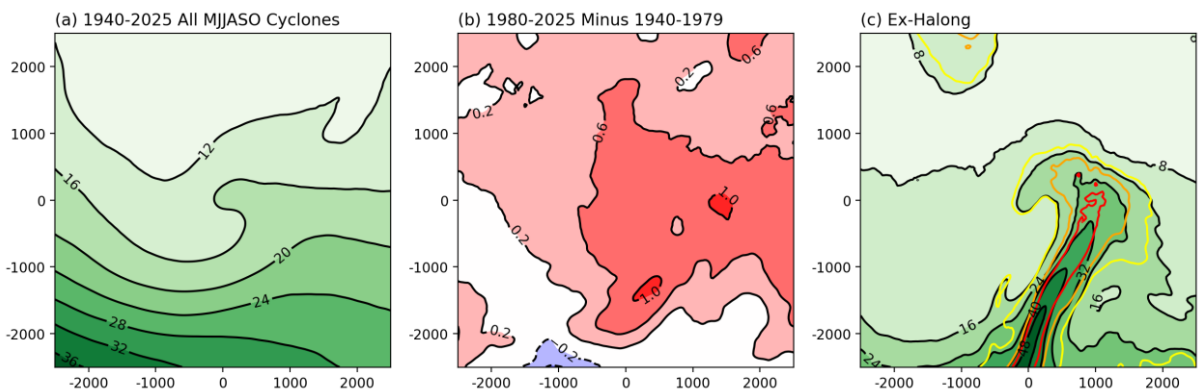

Figure 10. Storm-centric composite mean of total column water vapor (TCWV, shading in kg m$^{-2}$) for warm-season NPCs during 1940-2025 when located within (170°-200°E, 55°-65°N); (b) The mean difference of TCWV between NPCs in 1940-1979 and 1980-2025; (c) composite mean of ex-Halong during its occurrence within 170°-200°E, 55°-65°N (i.e., 0000-1800 UTC 12 October). In (c), the yellow, orange, and red contours enclose areas where ex-Halong's composite mean TCWV exceeds the 50th, 75th, and 90th percentiles, respectively, of all NPCs included in (a). Axes represent the zonal (x) and meridional (y) distance from the cyclone center in km.

Next, we examine ACTs over the North Pacific sector. Their long-term mean seasonality exhibits a clear bi-modal distribution (Fig. 11), with a primary peak in September (~one ACT every four years) and a secondary peak in February (~one ACT every seven years). Given that the long-term mean frequency of ACTs in October is less than one per ten years, ex-Halong can be regarded as a very rare event. Consistent with Fig. 9, Fig. 11 shows that ACTs have occurred more frequently in recent decades. In addition, the primary annual peak has shifted from February in the earlier period to September in the later period. The monthly frequency of ACTs has increased by 3-5 times in August and September, with a frequency exceeding one in three years in September of the more recent period. Since sea ice extent is lower in early fall than in early spring, the shift in ACT seasonality has likely contributed to stronger coastal impacts as discussed in the introduction. Whether the frequency and seasonality changes of ACTs are due to climate change or internal multidecadal variability of the climate merits further investigation.

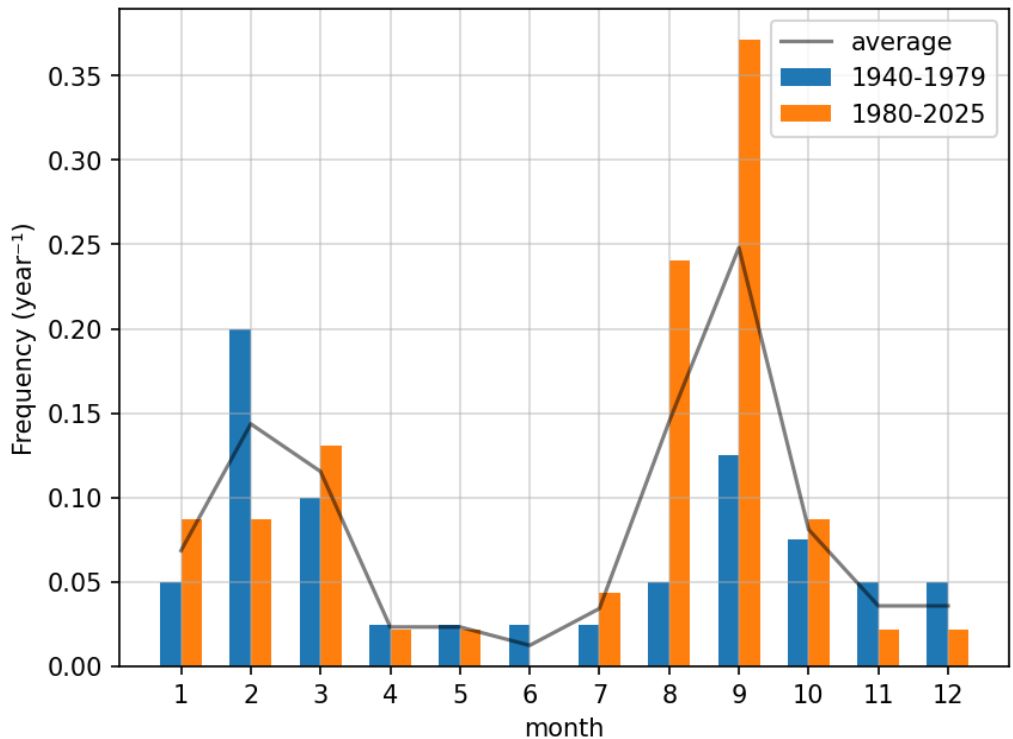

Figure 11. Monthly mean frequency of ACT events (gray line) per year. Blue and orange bars show the ACT frequencies per year before and after 1980, respectively.

Ex-Halong is further compared with other NPCs and ACTs in Fig. 12. Compared to the general population of warm-season cyclones affecting the Alaska-Bering Strait sector (i.e., NPCs; Figs. 12a-b), ex-Halong is exceptional in intensification rate and peak intensity, ranking within the top $2^{nd}$ and $3^{rd}$ percentiles, respectively, and its peak intensity occurred more poleward than the majority of NPCs ($28^{th}$ percentile). Ex-Halong is also an extreme case when compared specifically to the warm-season ACTs (Figs. 12c-d). Ex-Halong ranks in the $21^{st}$ percentile for lifetime maximum intensity among ACTs. Most remarkably, its extratropical re-intensification is particularly strong ($4^{th}$ highest percentile), and its latitude of maximum intensity falls in the $9^{th}$ highest percentile.

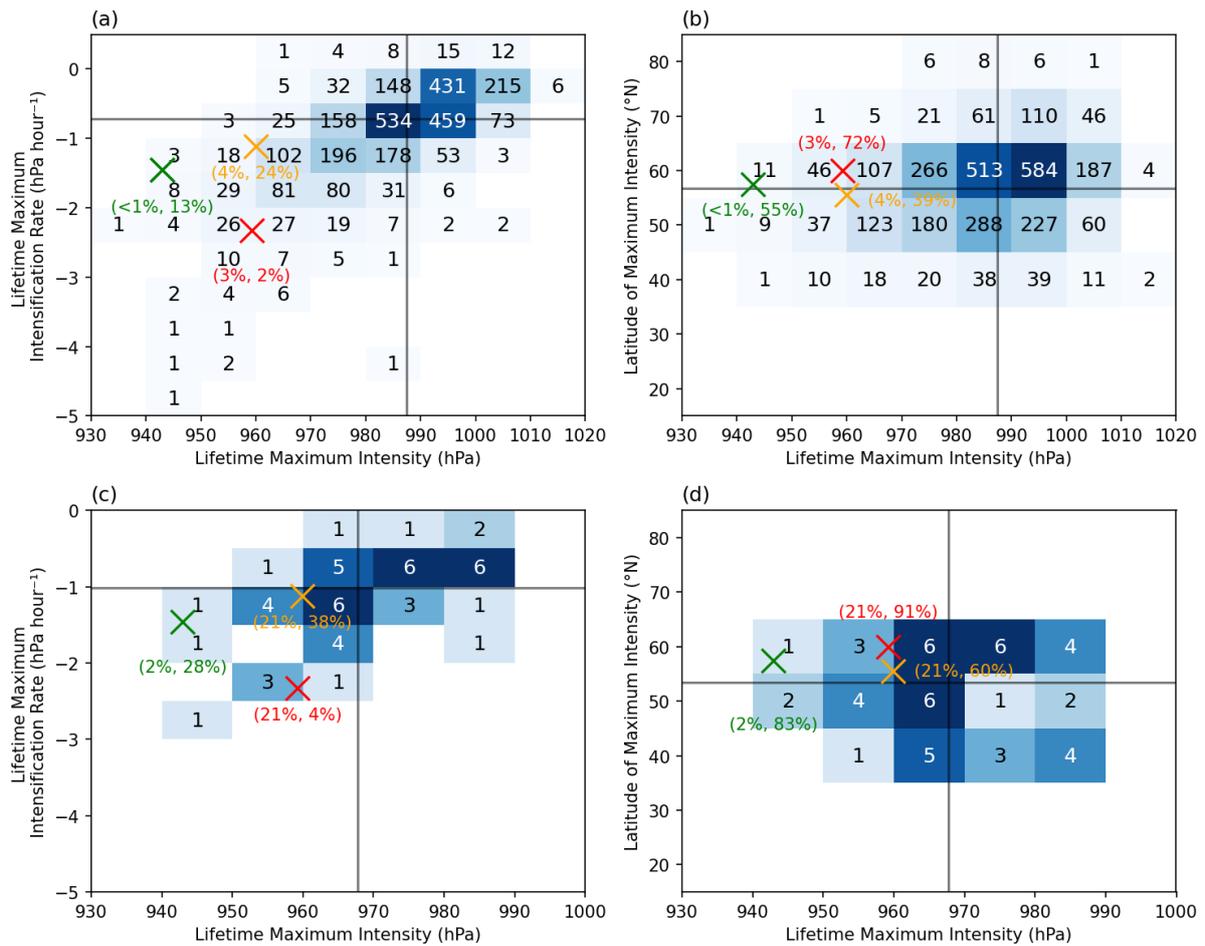

Figure 12. Two-dimensional histograms of SLP-based lifetime maximum intensity, maximum intensification rate, and latitude of the peak intensity for (a, b) warm-season NPCs and (c, d) warm-season ACTs. Red crosses mark the position of ex-Halong in the histograms, with red numbers indicating its percentile ranks within the corresponding cyclone group. Orange and green crosses and numbers indicate ex-typhoons Ampil in August 2024, and Merbok in September 2022, respectively. Gray lines represent sample medians. Intensity statistics are derived using SLP within the extratropic segments (latitude > 35°N) of the cyclone trajectories.

Ex-typhoon Ampil in August 2024 and ex-typhoon Merbok in September 2022 are also shown in Fig. 12. Like ex-Halong, both systems underwent (re-)intensification during their extratropical phases. Ex-typhoon Ampil attained a comparable peak intensity to ex-Halong but exhibited a maximum intensification rate that was only about half as large, and it reached its minimum SLP at lower latitudes (~55°N). Ex-Merbok was more extreme in terms of intensity than Ex-Halong, and it also produced strong surface wind, storm surges, and flooding in western Alaska. Among the ex-typhoons, ex-Halong stands out due to its unusually rapid re-intensification and its attainment of peak intensity at relatively high latitudes close to Alaska, highlighting the potential risks associated with suddenly intensifying extratropical cyclones near landfall. Its occurrence in October, near the end of the typhoon season, further underscores its unexpected nature compared to other recent ex-typhoons occurring in August and September.

## 8. Summary

Ex-Typhoon Halong (October 2025) was a recent example of tropical cyclones undergoing extratropical transition and subsequent re-intensification at high latitudes, producing severe coastal impacts in western Alaska. Utilizing both reanalysis data and numerical model simulations, this study investigates the re-intensification of ex-Halong and environmental processes affecting its evolution in the extratropics.

Ex-Halong transitioned from a symmetric warm-core typhoon into a strongly baroclinic, cold-core cyclone on 10 October. Its subsequent poleward propagation was strongly influenced by a pre-existing extratropical cyclone to the northwest, which shaped the large-scale flow pattern and steered ex-Halong into a favorable region for the upper-level forcing. Two re-intensification episodes occurred while ex-Halong propagated towards Bering Strait and Alaska. The first deepening episode was closely tied to diabatic heating facilitated by anomalously warm SSTs, while strong quasi-geostrophic ascent associated with vorticity advection by an upper-level jet, combined with diabatic heating, contributed to the second, more vigorous re-intensification episode.

WRF sensitivity experiments confirm the critical role of SST anomalies during Halong's re-intensification. The experiments showed that SST anomalies and the background SST trend both contribute to increased surface heat fluxes. The enhanced surface heat fluxes precondition ex-Halong's re-intensification by enhancing the tropospheric humidity and reducing static stability. Furthermore, the simulations also demonstrate that reliable track forecasts are closely related to the intensity forecasts of ex-Halong and also the nearby extratropical cyclone owing to their binary interaction.

From a climatological perspective, ex-Halong stands out as an exceptional case in terms of its intensity and intensification rate. Originating at an unusually low latitude, the system exhibited strong poleward moisture transport and an extratropical intensification rate ranking in the 4$^{th}$ highest percentile of warm-season ACTs and in the 2$^{nd}$ highest percentile of warm-season NPCs. Additionally, we show that ACTs have occurred more frequently in recent decades, and the shift of their annual peak frequency from February to September also implies stronger coastal impacts given the annual low sea ice extent in September. These factors underscore the growing risk and highlight the need for further research on severe, tropical-origin weather disruptions in Alaska in a changing climate. Given the role of SST in cyclone intensification and the destructive impacts of cyclones through storm surge, coupled models may be useful tools in assessing future changes in intense North Pacific cyclones.

*Acknowledgments.*

We acknowledge access to the ERA5 reanalysis via the Geoscience Data Exchange through NCAR's CISL, sponsored by the NSF. ZW and MY acknowledge the funding support by the Office of Naval Research through Grant N000141812216. JDD acknowledges the support of the NRL Base Program, Air-Sea-Ice Interactions and Intense Polar Lows, PE-0601153N, as well as the support of the Office of Naval Research PE-0602435N. AKD was supported by the NSF National Center for Atmospheric Research, a major facility sponsored by the U.S. National Science Foundation. Any opinions, findings and conclusions or recommendations expressed in this material do not necessarily reflect the views of NSF.

*Data Availability Statement.*

The ERA5 reanalysis data used in this study are publicly available through the NCAR Geoscience Data Exchange (https://gdex.ucar.edu/datasets/d633000) and the Copernicus Climate Data Store (https://cds.climate.copernicus.eu/). The Weather Research and Forecasting (WRF) model is an open-source community model available via GitHub (https://github.com/wrf-model/WRF). The cyclone tracking algorithm used to identify and track the systems in this study is available at https://github.com/alexcrawford0927/cyclonetracking.

Kuo, Y.-H., S. Low-Nam, and R. J. Reed, 1991: Effects of Surface Energy Fluxes during the Early Development and Rapid Intensification Stages of Seven Explosive Cyclones in the Western Atlantic. *Mon. Wea. Rev.*, **119**, 457–476, https://doi.org/10.1175/1520-0493(1991)119%253C0457:EOSEFD%253E2.0.CO;2.

Mesquita, M. S., D. E. Atkinson, and K. I. Hodges, 2010: Characteristics and Variability of Storm Tracks in the North Pacific, Bering Sea, and Alaska*. *Journal of Climate*, **23**, 294–311, https://doi.org/10.1175/2009JCLI3019.1.

NCAR/UCAR,: WRF USER SUPPORT. Accessed 17 January 2026, https://www2.mmm.ucar.edu/wrf/users/physics/wrf_physics_suites.php.

Overeem, I., R. S. Anderson, C. W. Wobus, G. D. Clow, F. E. Urban, and N. Matell, 2011: Sea ice loss enhances wave action at the Arctic coast: SEA ICE LOSS ENHANCES EROSION. *Geophys. Res. Lett.*, **38**, n/a-n/a, https://doi.org/10.1029/2011GL048681.

Parkinson, C. L., 2022: Arctic sea ice coverage from 43 years of satellite passive-microwave observations. *Front. Remote Sens.*, **3**, 1021781, https://doi.org/10.3389/frsen.2022.1021781.

Redilla, K., S. T. Pearl, P. A. Bieniek, and J. E. Walsh, 2019: Wind Climatology for Alaska: Historical and Future. *ACS*, **09**, 683–702, https://doi.org/10.4236/acs.2019.94042.

Rolph, R. J., A. R. Mahoney, J. Walsh, and P. A. Loring, 2018: Impacts of a lengthening open water season on Alaskan coastal communities: deriving locally relevant indices from large-scale datasets and community observations. *The Cryosphere*, **12**, 1779–1790, https://doi.org/10.5194/tc-12-1779-2018.

Rosen, Y., 2025: Alaska typhoon victims' losses of traditional foods go beyond dollar values | Alaska Beacon. Accessed 25 January 2026, https://alaskabeacon.com/2025/11/03/alaska-typhoon-victims-losses-of-traditional-foods-go-beyond-dollar-values/?utm_source=chatgpt.com.

Sinclair, V. A., M. Rantanen, P. Haapanala, J. Räisänen, and H. Järvinen, 2020: The characteristics and structure of extra-tropical cyclones in a warmer climate. *Weather Clim. Dynam.*, **1**, 1–25, https://doi.org/10.5194/wcd-1-1-2020.

Skamarock, W. C., and Coauthors, 2019: *A Description of the Advanced Research WRF Model Version 4*. UCAR/NCAR, accessed 17 January 2026, https://doi.org/10.5065/1DFH-6P97.
31

SUPPORTING INFORMATION

Table S1. Physics configurations in the WRF ensemble sensitivity experiments. The top row shows configurations held fixed for all ensemble members.

| WRF Version | Land Surface | Radiation SW/LW | Fractional Sea Ice |
|---:|---:|---:|---:|
| 4.2.2 | unified Noah | RRTMG (radt=20) | True |
| **PBL** | **Cumulus** | **Microphysics** | **Surface Layer** |
| YSU | Kain-Fritsch (new Eta) | Morrison 2-moment | MM5 |
| YSU | Kain-Fritsch (new Eta) | WSM 6-class graupel | MM5 |
| YSU | Kain-Fritsch (new Eta) | Thompson graupel | MM5 |
| YSU | New Tiedtke | Morrison 2-moment | MM5 |
| YSU | New Tiedtke | WSM 6-class graupel | MM5 |
| YSU | New Tiedtke | Thompson graupel | MM5 |
| YSU | Tiedtke | Morrison 2-moment | MM5 |
| YSU | Tiedtke | WSM 6-class graupel | MM5 |
| YSU | Tiedtke | Thompson graupel | MM5 |
| MYJ (Eta) TKE | Kain-Fritsch (new Eta) | Morrison 2-moment | Monin-Obukhov |
| MYJ (Eta) TKE | Kain-Fritsch (new Eta) | WSM 6-class graupel | Monin-Obukhov |
| MYJ (Eta) TKE | Kain-Fritsch (new Eta) | Thompson graupel | Monin-Obukhov |
| MYJ (Eta) TKE | New Tiedtke | Morrison 2-moment | Monin-Obukhov |
| MYJ (Eta) TKE | New Tiedtke | WSM 6-class graupel | Monin-Obukhov |
| MYJ (Eta) TKE | New Tiedtke | Thompson graupel | Monin-Obukhov |
| MYJ (Eta) TKE | Tiedtke | Morrison 2-moment | Monin-Obukhov |
| MYJ (Eta) TKE | Tiedtke | WSM 6-class graupel | Monin-Obukhov |
| MYJ (Eta) TKE | Tiedtke | Thompson graupel | Monin-Obukhov |

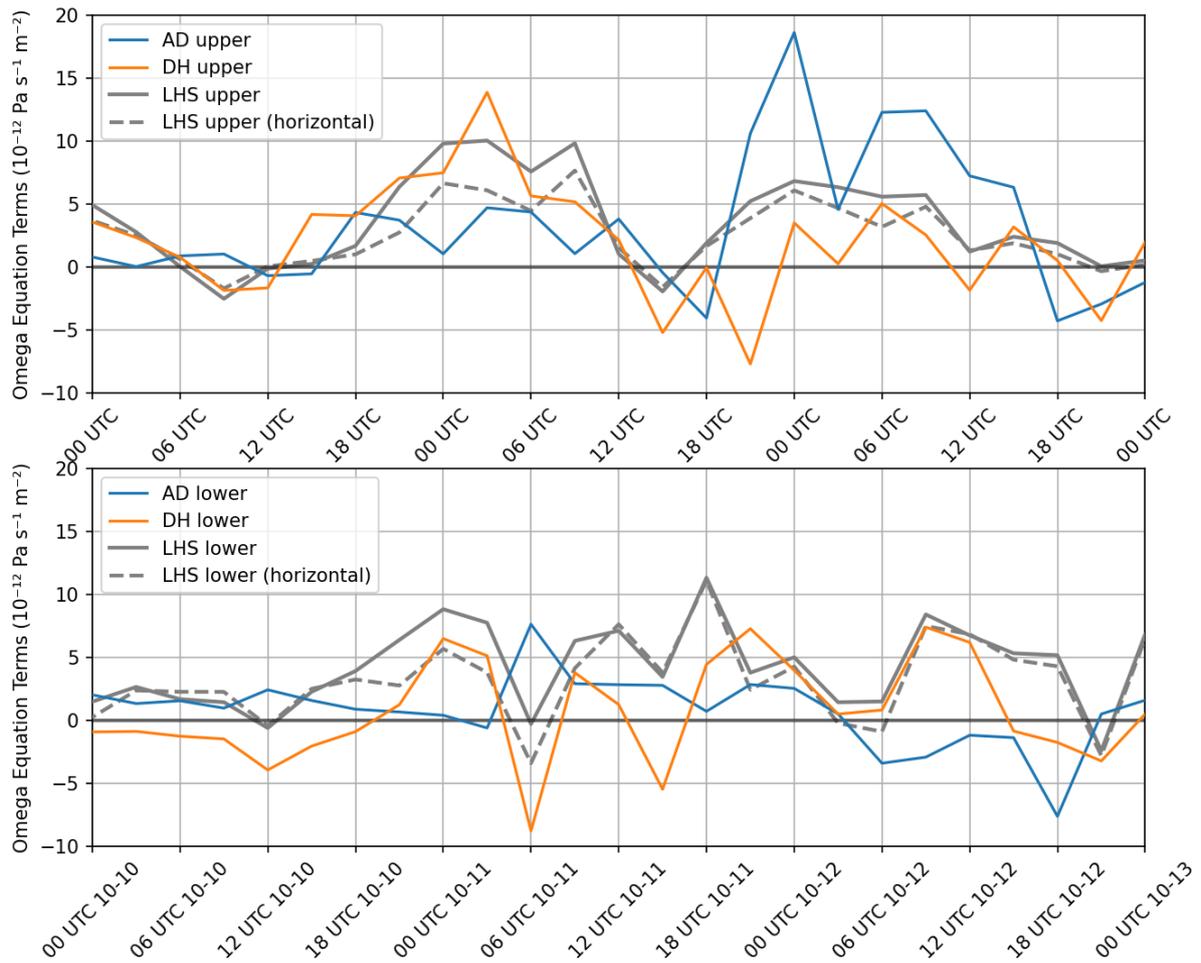

Figure S1. Temporal evolution of the LHS (gray; the dashed curve shows only horizontal component), the first RHS term (red; adiabatic forcing of Q vector convergence), and the third RHS term (orange; diabatic heating) of Eq. 5, averaged within a 500 km radius of the cyclone center for 600-300 hPa (top) and 900-600 hPa (bottom).

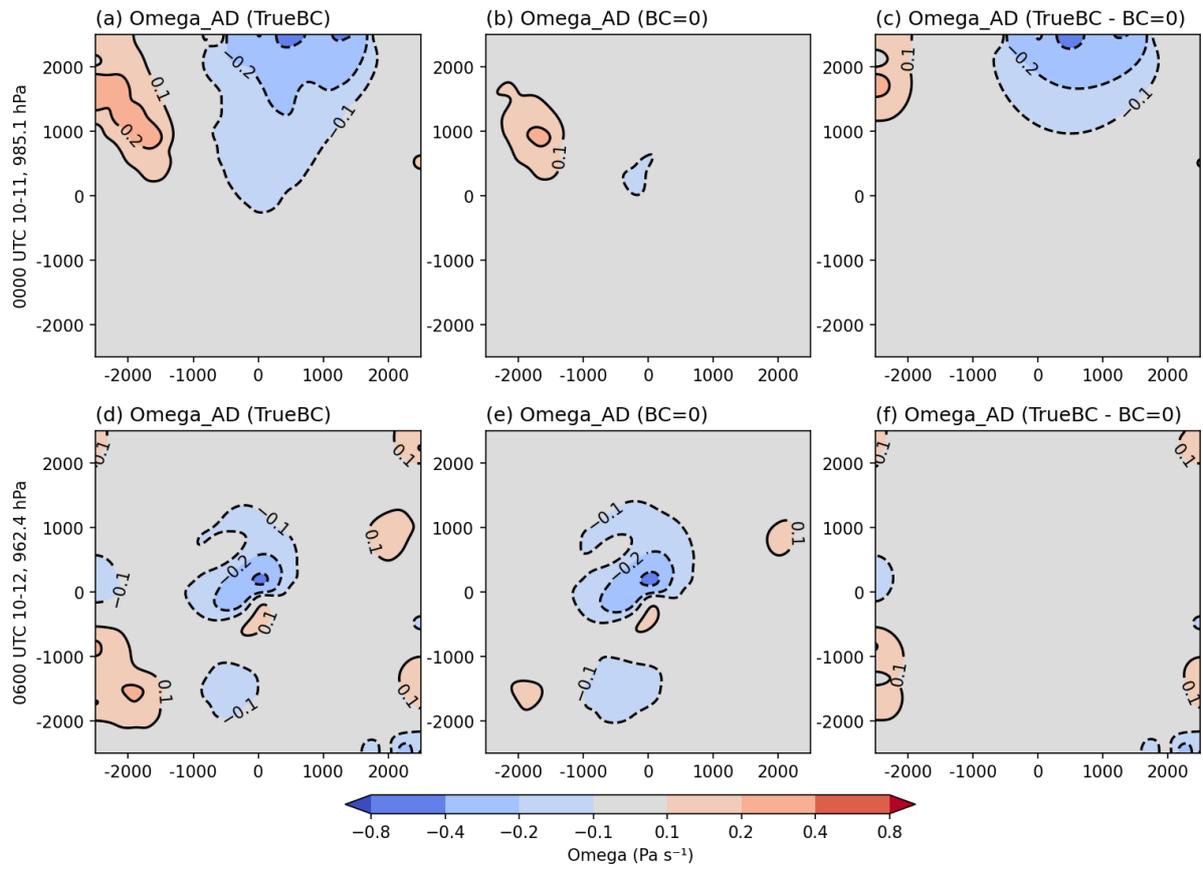

Figure S2. (Left) the adiabatic QG-derived omega using actual ERA5 boundary conditions and (middle) using Dirichlet boundary conditions, and (right) their differences. The two rows correspond to timesteps shown in Figure 4.

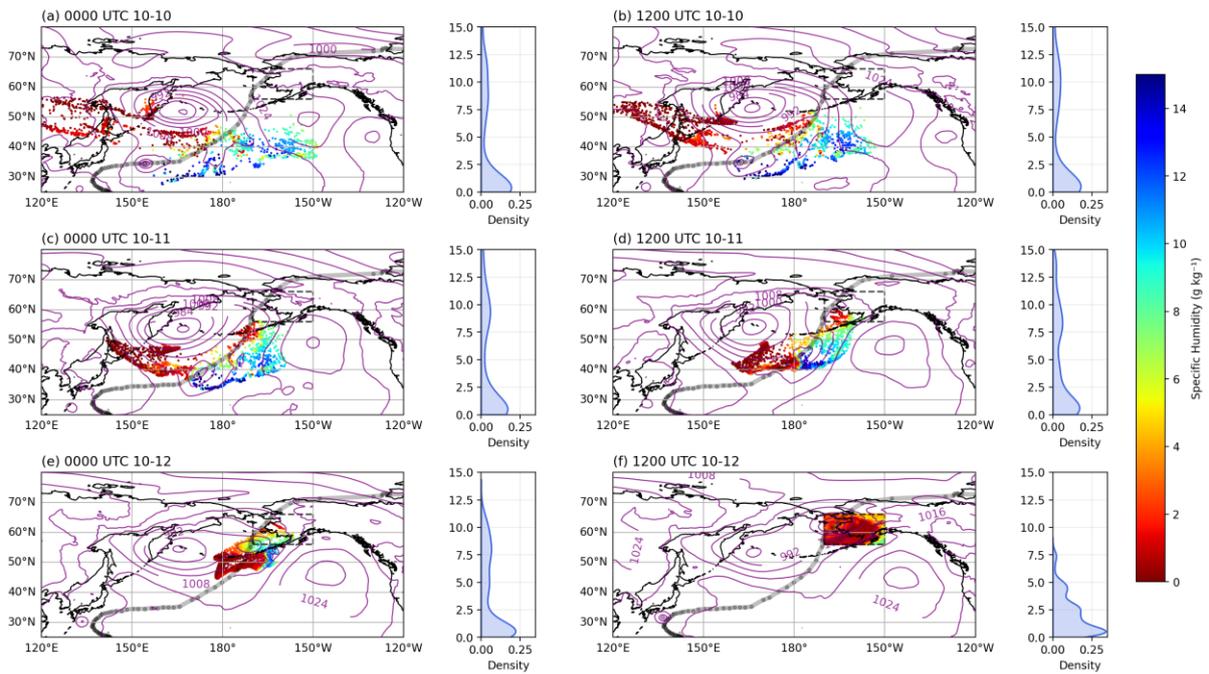

Figure S3. Parcel locations (colored dots) and specific humidity (g kg$^{-1}$) from backward trajectory analysis using the HYSPLIT model (Stein et al., 2015) with 3 hourly ERA5 reanalysis. 1,681 trajectories are initialized at 1200 UTC 12 October, 3000 m above the surface, spanning 56°-66°N at 0.25° intervals and 190°-210°E at 0.5° intervals (indicated by a black dashed box). The trajectories extend backward for 60 hours to 0000 UTC on 10 October. Gray curve shows the track of Halong, purple contours show SLP in hPa every 8 hPa. Side panels show the distribution of specific humidity.

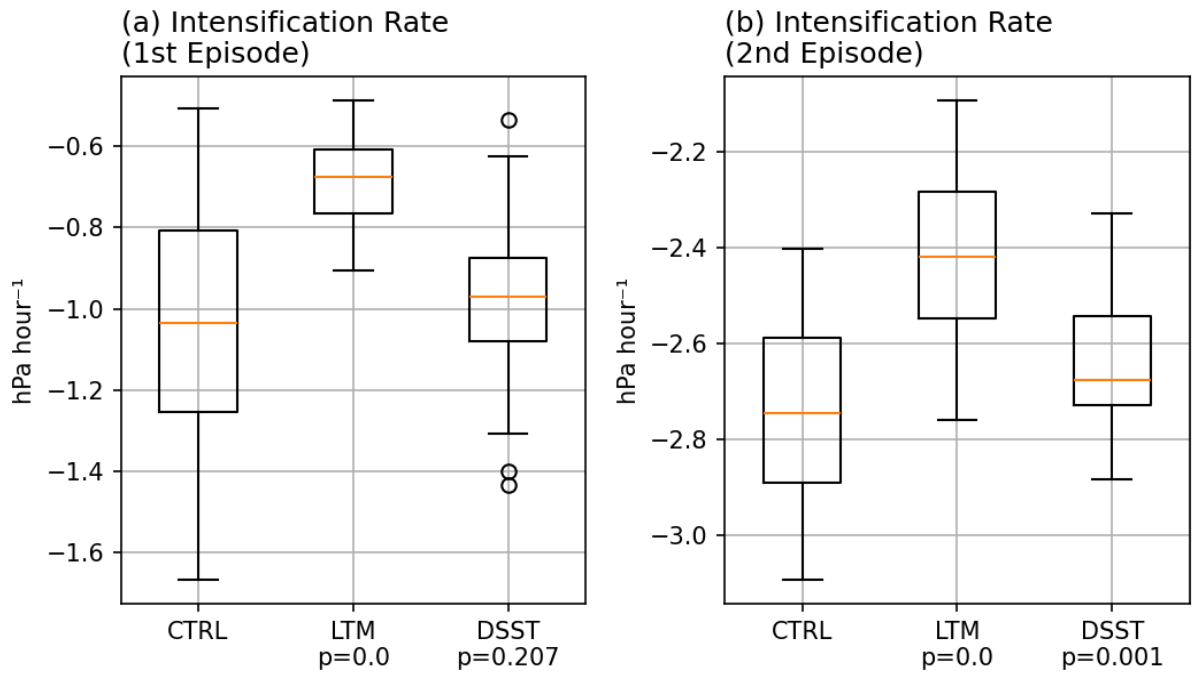

Figure S4 Same as Fig. 6f but for the strongest intensification rates during (a) the first re-intensification episode and (b) second re-intensification episode, in the WRF simulations.

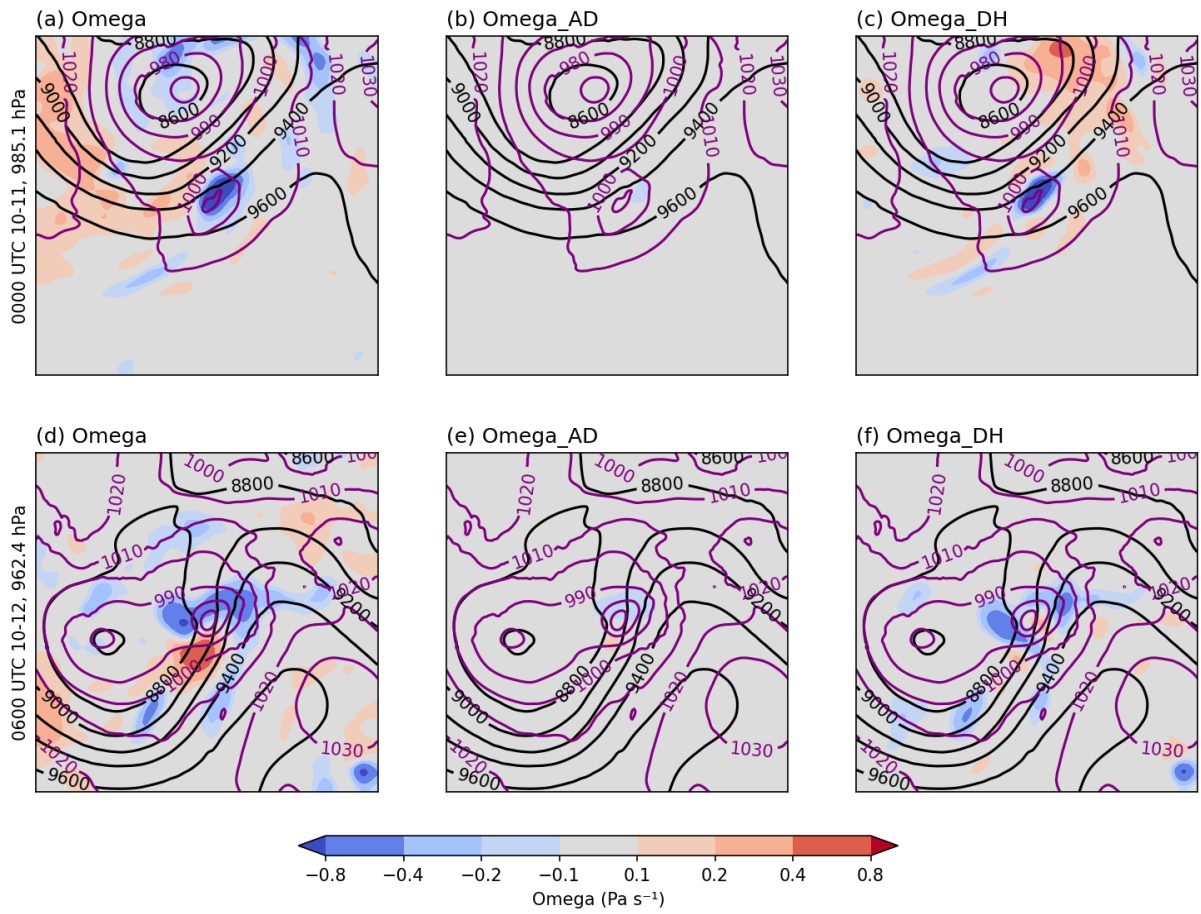

Figure S5. As in Figure 4, but showing omega fields averaged vertically between 900 and 600 hPa.

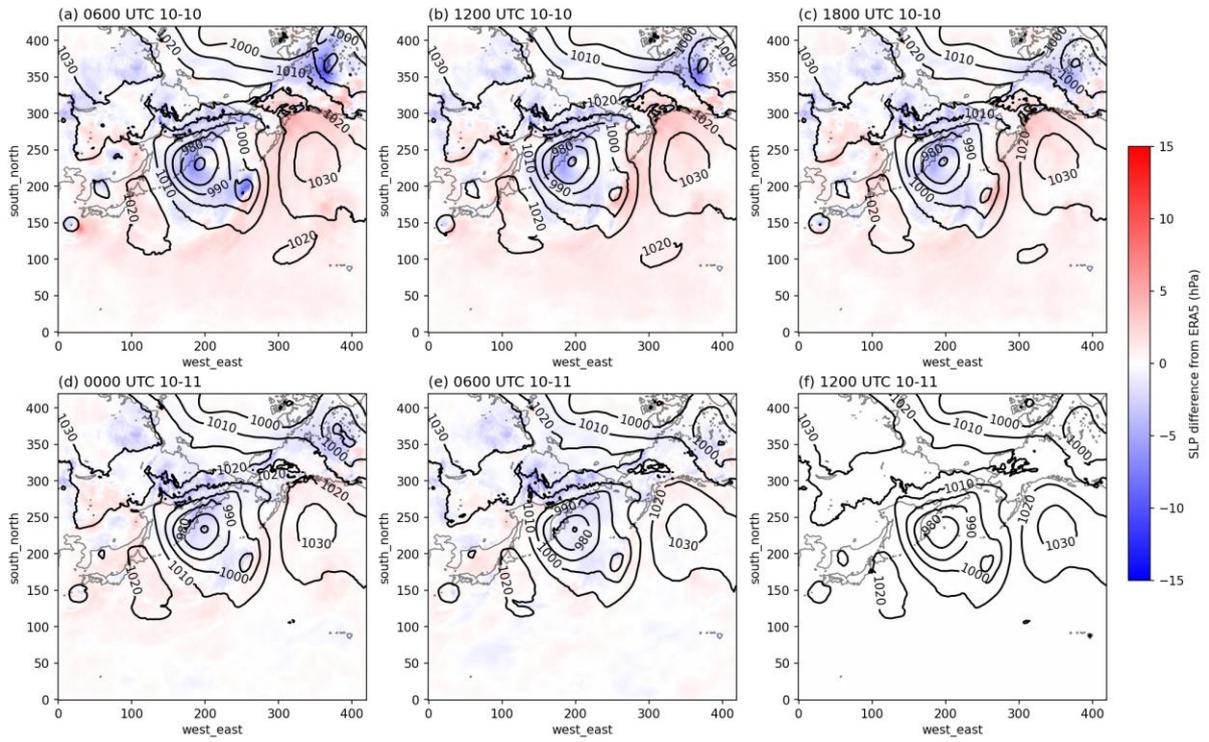

Figure S6. The SLP at 1200 UTC 11 October (black contours in hPa) and the SLP difference from ERA5 (shading in hPa), for simulations initialized in 6-hour increments from 0000 UTC 10 October.